\documentstyle{mn}
\input{epsf.tex}
\newif\ifAMStwofonts

\ifoldfss
  \ifCUPmtlplainloaded \else
    \NewTextAlphabet{textbfit} {cmbxti10} {}
    \NewTextAlphabet{textbfss} {cmssbx10} {}
    \NewMathAlphabet{mathbfit} {cmbxti10} {} 
    \NewMathAlphabet{mathbfss} {cmssbx10} {} 
  \fi  %
  \ifAMStwofonts
    \ifCUPmtlplainloaded \else
      \NewSymbolFont{upmath} {eurm10}
      \NewSymbolFont{AMSa} {msam10}
      \NewMathSymbol{\upi}     {0}{upmath}{19}
      \NewMathSymbol{\umu}     {0}{upmath}{16}
      \NewMathSymbol{\upartial}{0}{upmath}{40}
      \NewMathSymbol{\leqslant}{3}{AMSa}{36}
      \NewMathSymbol{\geqslant}{3}{AMSa}{3E}

       \let\le=\leqslant
       
    \fi
  \fi
\fi 

\ifnfssone
  \newmathalphabet{\mathit}
  \addtoversion{normal}{\mathit}{cmr}{m}{it}
  \addtoversion{bold}{\mathit}{cmr}{bx}{it}
  \newmathalphabet{\mathbfit} 
  \addtoversion{normal}{\mathbfit}{cmr}{bx}{it}
  \addtoversion{bold}{\mathbfit}{cmr}{bx}{it}
  \newmathalphabet{\mathbfss} 
  \addtoversion{normal}{\mathbfss}{cmss}{bx}{n}
  \addtoversion{bold}{\mathbfss}{cmss}{bx}{n}
  \ifAMStwofonts
    \ifCUPmtlplainloaded \else
      \UseAMStwoboldmath
      \makeatletter
      \new@mathgroup\upmath@group
      \define@mathgroup\mv@normal\upmath@group{eur}{m}{n}
      \define@mathgroup\mv@bold\upmath@group{eur}{b}{n}
      \edef\UPM{\hexnumber\upmath@group}
      \new@mathgroup\amsa@group
      \define@mathgroup\mv@normal\amsa@group{msa}{m}{n}
      \define@mathgroup\mv@bold\amsa@group{msa}{m}{n}
      \edef\AMSa{\hexnumber\amsa@group}
      \makeatother
      \mathchardef\upi="0\UPM19
      \mathchardef\umu="0\UPM16
      \mathchardef\upartial="0\UPM40
      \mathchardef\leqslant="3\AMSa36
      \mathchardef\geqslant="3\AMSa3E

       \let\le=\leqslant

    \fi
  \fi
\fi 

\ifnfsstwo
  \DeclareMathAlphabet{\mathbfit}{OT1}{cmr}{bx}{it}
  \SetMathAlphabet\mathbfit{bold}{OT1}{cmr}{bx}{it}
  \DeclareMathAlphabet{\mathbfss}{OT1}{cmss}{bx}{n}
  \SetMathAlphabet\mathbfss{bold}{OT1}{cmss}{bx}{n}
  \ifAMStwofonts
    \ifCUPmtlplainloaded \else
      \DeclareSymbolFont{UPM}{U}{eur}{m}{n}
      \SetSymbolFont{UPM}{bold}{U}{eur}{b}{n}
      \DeclareSymbolFont{AMSa}{U}{msa}{m}{n}
      \DeclareMathSymbol{\upi}{0}{UPM}{"19}
      \DeclareMathSymbol{\umu}{0}{UPM}{"16}
      \DeclareMathSymbol{\upartial}{0}{UPM}{"40}
      \DeclareMathSymbol{\leqslant}{3}{AMSa}{"36}
      \DeclareMathSymbol{\geqslant}{3}{AMSa}{"3E}

       \let\le=\leqslant

    \fi
  \fi
\fi 

\ifCUPmtlplainloaded \else
  \ifAMStwofonts \else 
    \def\upi{\pi}
    \def\umu{\mu}
    \def\upartial{\partial}
  \fi
\fi

\title{Energy relaxation in galaxies induced by an external environment
and/or incoherent internal pulsations}

\author[H. E. Kandrup]
  {Henry E. Kandrup\thanks{E-mail: kandrup@astro.ufl.edu}\\
Department of Astronomy,
Department of Physics,
and Institute for Fundamental Theory, University of Florida, Gainesville,
          FL 32611, USA}

\date{Accepted 2001 \hskip 1in .
      Received 2000 \hskip 1in .}

\pagerange{\pageref{firstpage}--\pageref{lastpage}}
\pubyear{2001}

\begin{document}

\maketitle

\label{firstpage}

\begin{abstract}
This paper explores the phenomenon of energy relaxation for stars in a galaxy
embedded in a high density environment that is subjected continually to 
perturbations reflecting  the presence of other nearby galaxies and/or
 incoherent internal pulsations. The analysis is similar to earlier 
analyses of energy relaxation induced by binary encounters between nearby 
stars and between stars and giant molecular clouds in that the perturbations 
are idealised as a sum of near-random events which can be modeled as diffusion 
and dynamical friction. However, the analysis differs in one important respect:
because the time scale associated  with these perturbations need not be short 
compared with the characteristic dynamical time $t_{D}$ for stars in the
original galaxy, the diffusion process cannot be modeled as resulting from a
sequence of instantaneous kicks, {\it i.e., white} noise. Instead, the 
diffusion is modeled as resulting from random kicks of finite duration, 
{\it i.e., coloured} noise characterised by a nonzero autocorrelation time 
$t_{c}$. A detailed analysis of coloured noise generated by sampling an 
Ornstein-Uhlenbeck process leads to a simpling scaling in terms of $t_{c}$ and 
an effective diffusion constant $D$. Interpreting $D$ and $t_{c}$ following 
early work by Chandrasekhar (1941) (the `nearest neighbour approximation')
implies that, for realistic choices of parameter values, energy relaxation 
associated with an external environment and/or internal pulsations could be 
important on times short compared with the age of the Universe.

\end{abstract}

\begin{keywords}
galaxies: structure -- galaxies: kinematics and dynamics
\end{keywords}

\section{MOTIVATION}
Dating back to the pioneering work of Chandrasekhar in the 1940's ({\it cf}. 
Chandrasekhar  1943a), it has been recognised that discreteness effects
will induce changes in the energy (and any other collisionless invariants)
in a system of stars idealised in a first approximation as a collisionless
equilibrium. In particular, by modeling interactions between individual
stars as an incoherent sum of binary encounters (gravitational Rutherford
scattering), Chandrasekhar computed a characteristic relaxation time $t_{R}$ 
on which the energy of a typical star will change by a factor of order unity. 
For small systems like open clusters and globular clusters, $t_{R}$ is 
typically short compared with $t_{H}$, the age of the Universe, and this 
energy relaxation has significant observable consequences, {\it e.g.}, for the 
development of core-halo structures in globular clusters ({\it cf}. Binney \& 
Tremaine 1987).

By contrast, for systems as large as entire galaxies, $t_{R}$ is typically 
long compared with $t_{H}$, so that it has become customary ({\it cf}. Binney 
\& Tremaine 1987) to ignore interactions between individual stars. However, 
such interactions are not the only irregularities which can induce energy 
relaxation. For example, in the 1950's Spitzer and Schwarzschild (1951, 1953) 
argued that interactions of stars with gas clouds in the Galactic disc 
will tend to increase their random velocities and their distance from the 
galactic plane on time scales short compared with $t_{H}$, thus providing an 
explanation of the observed correlations between kinematics and spectral type 
for Population I stars ({\it cf}. Mihalas \& Binney 1981).

The basic physical idea underlying this earlier work is that interactions 
between stars and/or gas clouds can be idealised as a sum of instantaneous 
near-random events, resulting in a phase space diffusion where, at least in a 
first approximation, the root mean squared change in quantities like the energy
grows as the square root of time, {\it i.e.}, 
${\delta}E_{rms}{\;}{\propto}{\;}t^{1/2}$. In particular, in his classic
review article on `Stochastic Problems in Physics and Astronomy,' Chandrasekhar
(1943b) recast the problem in the language of nonequilibrium statistical 
mechanics, allowing for the combined effects of diffusion and dynamical 
friction, with the effects of diffusion being modeled as Gaussian white noise,
{\it i.e.}, a sequence of instantaneous random impulses.

The objective of this paper is to argue that there are at least two other 
sources of near-random irregularities which, under certain circumstances, can 
induce significant energy relaxation on time scales short compared with 
$t_{H}$. 
\par\noindent
(1) Real galaxies are all perturbed to a lesser or greater extent by other 
nearby galaxies and, under appropriate circumstances, one might expect that 
these perturbations can be approximated as a superposition of near-random 
interactions of finite duration, {\it i.e.,} coloured noise. If the galaxy 
be situated in a comparatively
low density environment, the dominant sources of irregularities may be a small
number of satellite galaxies or, perhaps, a single large neighbour which 
execute near-periodic motions relative to the original galaxy. In this case,
it is clearly inappropriate to approximate the perturbations as random events.
However, a galaxy embedded in a high density rich cluster will be exposed to
a more irregular set of perturbations, associated over the course of time
with a larger number of different galaxies, and it would not seem unreasonable
to suppose that these perturbations could be modeled as an incoherent sum
of near-random interactions.
\par\noindent
(2) There is also growing evidence, both numerical and observational, that real
galaxies can exhibit finite amplitude oscillations which persist for times
long compared with a characteristic dynamical time, $t_{D}$. If these 
oscillations are dominated by a small number of normal or pseudo-normal modes,
they should be well approximated as nearly periodic perturbations. If, however,
the oscillations involve a comparatively large number of different modes, it 
would seem that their effects would be better modeled as a sum of near-random
perturbations, involving a superposition of a large number of different 
frequencies. A collection of different forces characterised by a random 
combination of frequencies combined with random phases is equivalent 
mathematically to (in general) coloured noise ({\it cf}. van Kampen 1981), 
with a finite autocorrelation time.

Section 2 begins by formulating the problem of energy relaxation abstractly in 
terms of $D$, the diffusion constant, ${\eta}$, the coefficient of dynamical 
friction, and $t_{c}$, the autocorrelation time, {\it i.e.,} the characteristic
time scale on which the random irregularities change. These quantities are 
then related to (1) the physical properties of a rich cluster, by assuming 
that the random perturbations acting on a given galaxy at given instant are 
associated with the effects of one or two particularly proximate galaxies 
(Chandrasekhar's [1941] `nearest neighbour approximation'); and (2) the 
effects of incoherent internal pulsations, by exploiting simple dimensional 
arguments. 
Section 3 summarises a numerical investigation of energy relaxation in which
diffusion is modeled as coloured noise sampling an Ornstein-Uhlenbeck process,
demonstrating the existence of a simple scaling in terms of $D$ and $t_{c}$.
Section 4 concludes by showing that, for reasonable choices of parameter 
values, energy relaxation induced by external perturbations and/or incoherent
internal oscillations can be significant over the age of the Universe, and
then speculating on possible implications for real galaxies.

\section{THEORETICAL CONSIDERATIONS}
\subsection{The abstract problem}
Begin by formulating the problem of energy relaxation abstractly for an
ensemble of point masses moving in a time-independent potential $V({\bf r})$, 
each of which is also subjected to dynamical friction and coloured noise, 
{\it i.e.,} random kicks of finite duration. This leads naturally to the 
consideration of a Langevin equation of the form ({\it cf}. van Kampen 1981)
\begin{equation}
{d^{2}x^{a}\over dt^{2}}=-{{\partial}V({\bf x})\over {\partial}x^{a}}
-{\eta}v^{a}+F^{a}, \qquad (a=x,y,z)
\end{equation}
where ${\eta}$ is the coefficient of dynamical friction and ${\bf F}$ is the
random force, which is treated as a stochastic variable. Assuming in the usual 
fashion that ${\bf F}$ corresponds to homogeneous Gaussian noise, its 
statistical properties are characterised completely by its first two moments, 
namely 
\begin{displaymath}
{\langle}F_{a}(t){\rangle}=0 \qquad {\rm and}
\end{displaymath}
\begin{equation}
{\langle}F_{a}(t_{1})F_{b}(t_{2}){\rangle}=
{\delta}_{ab}\;K(t_{1}-t_{2}),
\hskip .2in (a,b=x,y,z).
\end{equation}

For the special case of white noise, seemingly appropriate, {\it e.g.}, for 
modeling gravitational Rutherford scattering of nearby stars, the 
autocorrelation function $K$ is proportional to 
a Dirac delta, so that
\begin{equation}
K({\tau})=2{\eta}{\Theta}{\delta}_{D}({\tau}).{\;}
\end{equation}
Here the normalisation ensures that the friction and noise are related
by a Fluctuation-Dissipation Theorem in terms of a `temperature' ${\Theta}$. 
It is well known that such a Langevin description is equivalent ({\it cf}. 
Riskin 1989) to a Fokker-Planck equation of the form
\begin{equation}
{{\partial}f\over {\partial}t}+
{\bf v}{\cdot}{{\partial}f\over {\partial}{\bf x}}-
{{\partial}V({\bf x})\over {\partial}{\bf x}}{\cdot}
{{\partial}f\over {\partial}{\bf v}}=
-{{\partial}\over {\partial}{\bf v}}{\cdot}({\eta}{\bf v}f)+D
{{\partial}^{2}f\over {\partial}{\bf v}^{2}} ,
\end{equation}
where
\begin{equation}
D{\;}{\equiv}{\;}\int_{-\infty}^{\infty}\,d{\tau}K({\tau})=2{\Theta}{\eta}
\end{equation}
is the diffusion constant.

For coloured noise, the forces at nearby instants of time remain correlated
so that $K$ is no longer proportional to a Dirac delta. As a simple example, 
one can consider the so-called Ornstein-Uhlenbeck process ({\it cf}. Uhlenbeck 
\& Ornstein 1930, van Kampen 1981), 
for which $K$ decreases exponentially in time. Here
\begin{equation}
K({\tau})={{\eta}{\Theta}\over t_{c}}\,\exp(-|{\tau}|/t_{c}),
\end{equation}
where the autocorrelation time $t_{c}$ sets the time scale on which the
fluctuating random forces ${\bf F}$ change appreciably. The normalisation 
entering into this equation ensures that the effective diffusion constant
$D$, again defined by eq. (5), is fixed completely by ${\eta}$ and ${\Theta}$.

\subsection{Intuitive expectations}
The effects of energy diffusion induced by friction and white noise can be
derived directly from the Fokker-Planck description associated with the
Langevin equation (1). As shown, {\it e.g.,} in Habib, Kandrup, \& Mahon 
(1997), the Fokker-Planck equation (4) implies that the mean squared change in 
energy associated with multiple noisy integrations of the same initial 
condition satisfies
\begin{equation}
{d{\langle}({\delta}E)^{2}{\rangle}\over dt}=D{\langle}v^{2}{\rangle}
+{\eta}({\langle}v^{2}{\rangle}^{2}-{\langle}v^{4}{\rangle})
+{\eta}({\langle}v^{2}{\rangle}{\langle}x^{2}{\rangle}
-{\langle}v^{2}x^{2}{\rangle}).
\end{equation}
However, at early times one can approximate 
${\langle}v^{4}{\rangle}{\;}{\approx}{\;}{\langle}v^{2}{\rangle}^{2}$
and
${\langle}v^{2}x^{2}{\rangle}{\;}{\approx}{\;}{\langle}v^{2}{\rangle}
{\langle}x^{2}{\rangle}$, so that
\begin{equation}
{d{\langle}({\delta}E)^{2}{\rangle}\over dt}{\;}{\approx}{\;}
D{\langle}v^{2}{\rangle}.
\end{equation}
Noting further that ${\langle}v^{2}{\rangle}{\;}{\sim}{\;}|E_{0}|$, where 
$E_{0}$ is the initial energy, one then infers that
\begin{equation}
{d{\langle}({\delta}E)^{2}{\rangle}\over dt}{\;}{\sim}{\;}D|E_{0}|,
\end{equation}
which implies a fractional root mean squared change in energy
\begin{equation}
{{\delta}E_{rms}\over |E_{0}|}{\;}{\;}{\sim}{\;}\left( 
{Dt\over |E_{0}|}\right)^{1/2}.
\end{equation}
Given that an ensemble of different initial conditions each evolved in the
presence of friction and noise can be viewed as a sum of random samplings
of Langevin simulations with different initial conditions, each satisfying 
eq. (10), it is clear that
this relation should also hold quite generally for any ensemble of orbits with
the same initial energy. The validity of this relation has been established 
numerically for a variety of different two- and three-dimensional potentials 
({\it cf}. Habib, Kandrup, \& Mahon 1997).

The obvious question is: how will this result be changed if one considers
coloured noise with a finite autocorrelation time? Recent work indicates 
that coloured noise impacts orbits via a resonant coupling between
the natural frequencies associated with the perturbation and the natural
frequencies of the orbits (Pogorelov \& Kandrup 1999, Kandrup, Pogorelov, \&
Sideris 2000). White noise is characterised by a flat spectral density with
power at all frequencies (the square of the spectral density is the Fourier 
transform of
the autocorrelation function $K$), so that it can couple to more or less 
anything. Replacing white noise by coloured Ornstein-Uhlenbeck noise with a 
nonzero autocorrelation time yields instead a spectral density 
\begin{equation}
S({\omega}){\;}{\propto}{\;}{{\alpha}^{2}\over {\omega}^{2}+{\alpha}^{2}},
\end{equation} 
with ${\alpha}=t_{c}^{-1}$, which cuts off for ${\omega}{\;}
{\gg}{\;}{\alpha}$. Given that the characteristic frequencies associated with
the unperturbed orbits typically scale as $t_{D}^{-1}$, with $t_{D}$ a
characteristic dynamical time, one might therefore expect (i) that coloured
noise with $t_{c}{\;}{\ll}{\;}t_{D}$ will have virtually the same effect as 
white noise; but (ii) that coloured noise with the same $D$ but
$t_{c}{\;}{\gg}{\;}t_{D}$ will be significantly less efficient than white 
noise in inducing energy relaxation. This in fact turns out to be the case. 
The numerical experiments described in the following section lead to the
simple scaling
\begin{equation}
{{\delta}E_{rms}\over |E|}{\;}{\sim}{\;} 
{\;}\left( { Dt\over |E| }  \right)^{1/2} \cases{
\;\;1 & for $t_{c}<t_{D}$\cr
 & \cr
\left( {t_{D}\over t_{c}} \right)  & for $t_{c}>t_{D}$.\cr}
\end{equation}

\subsection{Connection with physical parameters}
It remains to relate $D$, ${\eta}$, and $t_{c}$ to observable parameters 
characterising a real galaxy, considering first the effects of a high density 
environment. 

Begin by observing that, on dimensional grounds ({\it cf}. eq. [6]), the 
diffusion constant
\begin{equation}
D{\;}{\sim}{\;}F^{2}t_{c},
\end{equation}
where $F$ denotes the characteristic magnitude of the force per unit mass 
associated with the perturbing galaxies. Following Chandrasekhar (1941) or 
Chandrasekhar \& von Neumann (1942), suppose now that these effects are
associated primarily with one or two nearby galaxies. It then follows on
dimensional grounds that
\begin{equation}
F{\;}{\sim}{\;}{GM_{g}\over R_{s}^{2}}
\end{equation}
and
\begin{equation}
t_{c}{\;}{\sim}{\;}{R_{s}\over V_{s}},
\end{equation}
where $M_{g}$ represents the mass of a typical perturbing galaxy, $R_{s}$ the
typical separation between galaxies, and $V_{s}$ the typical relative velocity 
between galaxies. This implies that 
\begin{equation}
D{\;}{\sim}{\;}{G^{2}M_{g}^{2}\over R_{s}^{3}V_{s}}.
\end{equation}
However, by interpreting the `temperature' ${\Theta}$ as a characteristic 
squared velocity with which the perturbing galaxies move through space, one
has ${\Theta}{\;}{\sim}{\;}V_{s}^{2}$, so that
\begin{equation}
{\eta}{\;}{\sim}{\;}{G^{2}M_{g}^{2}\over R_{s}^{3}V_{s}^{3}}.
\end{equation}

It is easily seen that an application of a similar argument to stars moving
within a galaxy leads essentially to the standard expression for the 
coefficient of dynamical friction associated with binary encounters. 
Specifically, if $M_{g}$ is reinterpreted as the mass $m$ of a typical star and
$V_{s}$ as a characteristic relative velocity $V_{*}$ between nearby stars, and
$R_{s}$ is replaced by the typical separation ${\sim}{\;}n^{-1/3}$ 
between neighbouring stars, one concludes that 
\begin{equation}
{\eta}=t_{R}^{-1}{\;}{\sim}{\;}{G^{2}m^{2}n\over V_{*}^{3}},
\end{equation}
where $n$ represents a characteristic stellar density.
Equation (18) agrees ({\it cf}. Binney \& Tremaine 1987) with the standard 
expression for $t_{R}$, modulo the absence of the Coulomb logarithm 
$\ln {\Lambda}$ which would 
ordinarily appear in the numerator. This additional factor, which arises in a 
more careful analysis of discreteness effects, reflects the fact that weaker, 
more distant interactions also contribute to energy relaxation, thus reducing
the value of $t_{R}$ somewhat. In this sense, one infers that {\it an analysis 
of energy relaxation based on eqs. (15) - (17) will, if anything, 
underestimate the importance of environmental effects.}

Estimating the effects of internal pulsations seems more uncertain, but one
{\it can} proceed using dimensional analysis. Clearly one can write
\begin{equation}
F{\;}{\sim}{\;}{\alpha}{GM_{g}\over R_{g}^{2}}, \qquad
t_{c}{\;}{\sim}{\;}{\beta}t_{D}, \qquad {\rm and} \qquad
{\Theta}{\;}{\sim}{\;}{\gamma}V^{2},
\end{equation}
where ${\alpha}$, ${\beta}$, and ${\gamma}$ represent dimensionless constants.
Here the constant ${\alpha}$ essentially sets the characteristic amplitude of 
the irregularities associated with the internal pulsations as compared with 
the bulk force associated with the galaxy as a whole and, as such, should
be small compared with unity. One would thus anticipate that 
${\alpha}{\;}{\ll}{\;}1$. Appropriate values for the constants ${\beta}$ and 
${\gamma}$ are less clear, but it would at least seem reasonable to assume
that ${\beta}$ is not significantly larger than unity, {\i.e.,} the 
autocorrelation time $t_{c}$ is not much larger than the dynamical time 
$t_{D}$.
The characteristic size of ${\gamma}$, which enters into $D$ but {\it not} 
${\eta}$, is more difficult to estimate. Fortunately, however, it follows from 
the simulations described in Section 3 ({\it cf}. eq. [12]) that the rate of 
energy relaxation is 
fixed completely by $D$ and $t_{c}$, independent of ${\eta}$, so that its 
precise value is immaterial. In any event, given this scaling one infers that
\begin{equation}
D{\;}{\sim}{\;}
{\alpha}^{2}{\beta}{G^{2}M_{g}^{2}\over R_{g}^{4}}t_{D}{\;}{\sim}{\;}
{\alpha}^{2}{\beta}{G^{2}M_{g}^{2}\over R_{g}^{3}V_{*}},
\end{equation}
where one has used the fact that $t_{D}{\;}{\sim}{\;}R_{g}/V_{*}$.
\section{NUMERICAL EXPERIMENTS}
\subsection{What was computed}
Solutions to the Langevin equation (1) for a constant coefficient of
dynamical friction and coloured noise sampling the Ornstein-Uhlenbeck
process (7) were obtained for three different classes of potentials.
The most interesting astronomically, but by far the most expensive
computationally, were the triaxial generalisations of the Dehnen (1993)
potentials, which have been considered extensively by Merritt and
collaborators ({\it e.g.} Merritt \& Fridman 1996). These correspond to 
potentials generated self-consistently from the triaxial mass density
\begin{equation}
{\rho}(m)={(3-{\gamma})\over 4{\pi}abc}m^{-{\gamma}}(1+m)^{-(4-{\gamma})}
\end{equation}
with
\begin{equation}
m^{2}={x^{2}\over a^{2}}+{y^{2}\over b^{2}}+{z^{2}\over c^{2}},
\end{equation}
for various choices of axis ratios $a:b:c$ and cusp index ${\gamma}$.
Because of the cost associated with integrations of this potential, a
careful exploration of parameter space was prohibitive, so that more
systematic investigations were performed instead for the toy potentials
\begin{equation}
V(x,y,z)={1\over 2}(a^{2}x^{2}+b^{2}y^{2}+c^{2}z^{2})-
{M_{BH}\over \sqrt{r^{2}+{\epsilon}^{2}}},
\end{equation}
with $r^{2}=x^{2}+y^{2}+z^{2}$, which have been shown (Kandrup \& Sideris
2000) to reproduce much of the observed behaviour for orbit ensembles 
evolved in the triaxial Dehnen potentials. To test the genericity of the
basic conclusions, integrations were also performed for a completely different
class of computationally inexpensive potentials with
\begin{displaymath}
V(x,y,z)=-(x^{2}+y^{2}+z^{2})+{1\over 4}(x^{2}+y^{2}+z^{2})^{2}
\end{displaymath}
\begin{equation}
{\;\;\;\;\;\;\;\;\;\;\;\;\;\;\;}
-{1\over 4}(x^{2}y^{2}+ay^{2}z^{2}+bz^{2}x^{2}). 
\end{equation}
These potentials, which were explored systematically in Kandrup, Pogorelov,
\& Sideris (2000), constitute three-dimensional generalisations of the 
two-dimensional dihedral potential first considered by Armbruster, 
Guckenheimer, \& Kim (1989).

The experiments that were performed each involved selecting ensembles of $8000$
initial conditions, all with the same energy, and, for given choices of $\eta$,
${\Theta}$, and $t_{c}$, integrating these initial conditions into the future
for a time $t{\;}{\sim}{\;}100-200\,t_{D}$. In order to determine whether 
energy
relaxation impacts regular and chaotic orbits in the same fashion, regular and
chaotic ensembles were considered separately. For the potentials considered,
the typical particle speed $v$ and the dynamical time $t_{D}$ are both of 
order unity. For this reason, most of the experiments assumed that 
${\Theta}{\;}{\sim}{\;}1$. The coefficient of dynamical friction ${\eta}$,
which defines the diffusion constant $D=2{\Theta}{\eta}$, was allowed to vary
in the range $10^{-8}{\;}{\le}{\;}{\eta}{\;}{\le}{\;}10^{-2}$. The 
autocorrelation time $t_{c}$ was chosen to satisfy 
$10^{-2}{\;}{\le}{\;}t_{c}{\;}{\le}{\;}10^{3}$. It 
was verified explicitly that integrations with autocorrelation times as short 
as $t_{c}=0.01$ yield results essentially indistinguishable from integrations
with white noise.

White noise integrations were performed using a fixed time step algorithm 
developed by 
Griner, Strittmatter, \& Honerkamp (1988). Coloured noise integrations were
performed using an extension of this algorithm developed by I. V. Pogorelov
as part of his Ph.~D. thesis (see Pogorelov \& Kandrup 1999). The basic idea
underlying this extension is to (i) use a pseudo-random number generator to
generate white noise $X(t)$, and then (ii) transform $X(t)$ into coloured
noise $Y(t)$ sampling the Ornstein-Uhlenbeck process by solving the stochastic 
differential equation  
\begin{equation}
{dY\over dt}+{\alpha}Y=X(t),
\end{equation}
with ${\alpha}=t_{c}^{-1}$. 
Viewed as an initial value problem ({\it cf}. Chandrasekhar 1943b), eq. (25)
has the property that any nontrivial dependence on initial conditions dies 
away exponentially on a time scale ${\alpha}^{-1}$, so that, at late times, 
the random variable $Y(t)$ satisfies
\begin{equation}
{\langle}Y(t){\rangle}=0 \qquad {\rm and} \qquad
{\langle}Y(t_{1})Y(t_{2}{\rangle}=2\exp (-{\alpha}|t_{1}-t_{2}|).
\end{equation}
In other words, the late time solution to eq. (25) for $X(t)$ given as white
noise is coloured noise sampling the Ornstein-Uhlenbeck process $Y(t)$.
\subsection{What was found}
Perhaps the single most important conclusion derived from these simulations
is that the observed behaviour of different orbit ensembles is independent
of energy and choice of potential, {\it i.e.}, {\it the results appear to be
universal.} Modulo the numerical constant ${\cal A}$ entering into the 
relation ${\langle}v^{2}{\rangle}={\cal A}E$, which is exploited in the
derivation of eqs. (9) and (10), different ensembles yielded very similar 
results.
This is not surprising. When considering the problem of diffusion on a constant
energy hypersurface, different types of orbits and/or different potentials
can yield very different behaviour ({\it cf}. Siopis \& Kandrup 2000): Regular 
orbits are confined to three-dimensional phase space tori whereas chaotic
orbits move on a higher-dimensional phase space hypersurface which is impacted
by a complex Arnold web, the details of which can depend sensitively on the 
choice of potential and initial condition. However, energy relaxation involves
motion orthogonal to the hypersurfaces of constant energy, so that most of
these details are irrelevant. As discussed more carefully below, only for very 
long autocorrelation times $t_{c}$ does it appear that the qualitative results 
depend at all on the form of the potential or whether the orbits be regular
as opposed to chaotic.

It was also found that, as would be expected, coloured noise acts 
diffusively, so that the root mean squared energy for an orbit ensemble
grows as the square root of the integration time, {\it i.e.}, 
${\delta}E_{rms}{\;}{\propto}{\;}t^{1/2}$. This is, {\it e.g.,} evident from
FIG. 1, which exhibits data for an ensemble of 8000 chaotic orbits with $E=1$ 
evolved in the generalised dihedral potential for a total time $t=256$ with 
$a=b=1$, allowing for coloured noise with ${\Theta}=1$ for several different
choices of ${\eta}$ and $t_{c}$.

\begin{figure}
\centering
\centerline{
        \epsfxsize=8cm
        \epsffile{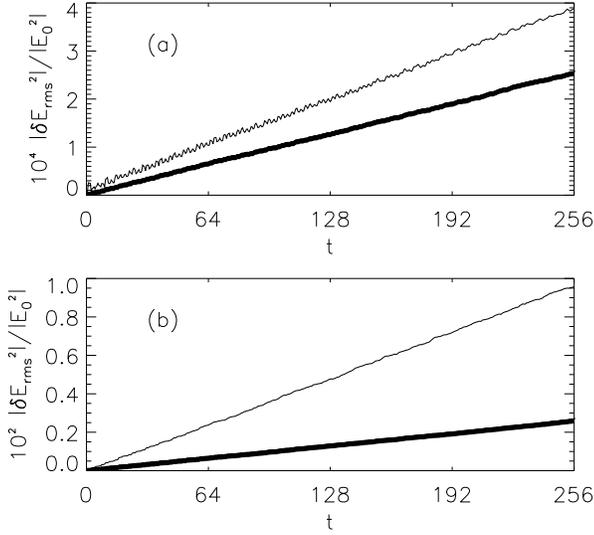}
           }
        \begin{minipage}{10cm}
        \end{minipage}
        \vskip -0.3in\hskip -0.0in
\caption{ 
$|{\delta}E_{rms}^{2}|/|E^{2}_{0}|$, the fractional change in squared energy 
for an 8000 orbit ensembles evolved for a time $t=256$ in the generalised
dihedral potential with $E=4$ and $a=b=1$ in the presence of coloured noise 
with ${\Theta}=1$ for several different choices of ${\eta}$ and $t_{c}$.
(a) The top curve has ${\eta}=10^{-4}$ and $t_{c}=10.0$. The lower has ${\eta}=
10^{-6}$ and $t_{c}=1.0$. (b) The top curve has ${\eta}=10^{-5}$ and 
$t_{c}=0.1$. The lower has  ${\eta}=10^{-5}$ and $t_{c}=1.0$.}
\vspace{0.0cm}
\label{landfig}
\end{figure}

\begin{figure}
\centering
\centerline{
        \epsfxsize=8cm
        \epsffile{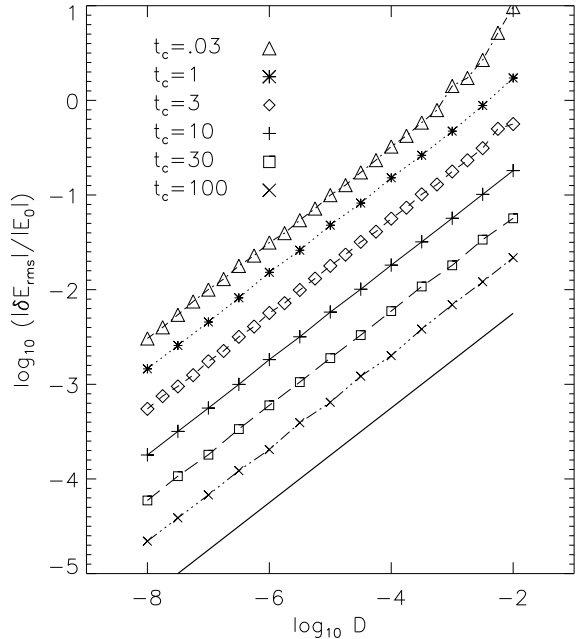}
           }
        \begin{minipage}{10cm}
        \end{minipage}
        \vskip -0.3in\hskip -0.0in
\caption{ 
(a) $|{\delta}E_{rms}|/|E_{0}|$, the fractional change in energy for 8000
orbit ensembles evolved in the dihedral potential with $E=4$ and $a=b=1$
after a time $t=256$, viewed as a function of $D$ for different choices of
$t_{c}$. The solid line has slope $0.5$.}
\vspace{0.0cm}
\label{landfig}
\end{figure}

Similarly, as for the case of white noise, one finds that the individual
values of ${\Theta}$ and ${\eta}$ are irrelevant. For fixed autocorrelation
time $t_{c}$, all that matters is the diffusion constant $D=2{\Theta}{\eta}$.
Moreover, as for the case of white noise, the amount of energy relaxation,
as probed by ${\delta}E_{rms}$, scales as $D^{1/2}$ for fixed $t_{c}$. This
fact is illustrated in FIG. 2 which, for the ensemble of initial conditions
used to generate FIG. 1, exhibits the fractional root mean squared change 
in energy, $|{\delta}E_{rms}|/|E_{0}|$, at time $t=256$ as a 
function of $D$ for various choices of $t_{c}$.

\begin{figure}
\centering
\centerline{
        \epsfxsize=8cm
        \epsffile{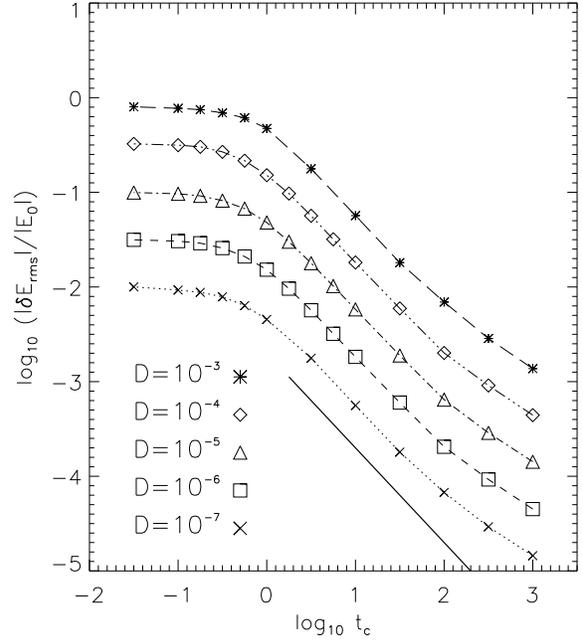}
           }
        \begin{minipage}{10cm}
        \end{minipage}
        \vskip -0.3in\hskip -0.0in
\caption{ 
(a) $|{\delta}E_{rms}|/|E_{0}|$, the fractional change in energy for 8000
orbit ensembles evolved in the dihedral potential with $E=4$ and $a=b=1$
after a time $t=256$, viewed as a function of $t_{c}$ for different choices of
$D$. The solid line has slope $-1.0$.}
\vspace{0.0cm}
\label{landfig}
\end{figure}

For fixed diffusion constant $D$, the dependence on $t_{c}$ is less trivial.
As expected, one discovers that, for $t_{c}{\;}{\ll}{\;}t_{D}$, the effects
of coloured noise are virtually identical to the effects of white noise, which
can be viewed as a singular limit of eq. (5) or (11). Alternatively, for 
$t_{c}{\;}{\gg}{\;}t_{D}$, the amount of energy relaxation for fixed $D$ is
significantly reduced. This is, {\it e.g.}, illustrated in FIG. 3, which, once
again for the ensemble used in FIG. 1, exhibits $|{\delta}E_{rms}|/|E_{0}|$ at 
time $t=256$ as a function of $t_{c}$ for various choices of $D$. It is evident
that, even for $t_{c}{\;}{\gg}{\;}t_{D}$, the slopes of the curves in this 
Figure are not completely constant, but it does appear that, for 
$0{\;}{\le}{\;}\log_{10}t_{c}{\;}{\le}{\;}2$, {\it i.e.,} for $t_{c}$ between
roughly $t_{D}$ and $100t_{D}$, the amount of energy relaxation scales
as $t_{c}^{-1}$. In this approximation, the results illustrated in FIGS. 1 - 3
yield the scaling summarised in eq. (12).

It is well known that, as far as energy relaxation is concerned, white noise 
impacts regular and chaotic orbits identically. This also appears to be the
case for coloured noise, at least for values of the autocorrelation time
$t_{c}<30t_{D}$ or so. Interestingly, however, there {\it are} distinct
differences for very large values of $t_{c}$. As illustrated in FIG. 4, it
appears that, for $t_{c}>30t_{D}$ or so, {\it regular orbits are impacted
more than chaotic orbits}. In particular, for very large $t_{c}$ the scaling
relation (12) drastically underestimates the importance of energy relaxation.
Why this is the case is not completely clear. Given, however, that coloured
noise acts via a resonant coupling to the orbits, it seems reasonable to
conjecture that this effect arises because regular orbits can 
couple more coherently to the slowly varying random forces.

\begin{figure}
\centering
\centerline{
        \epsfxsize=8cm
        \epsffile{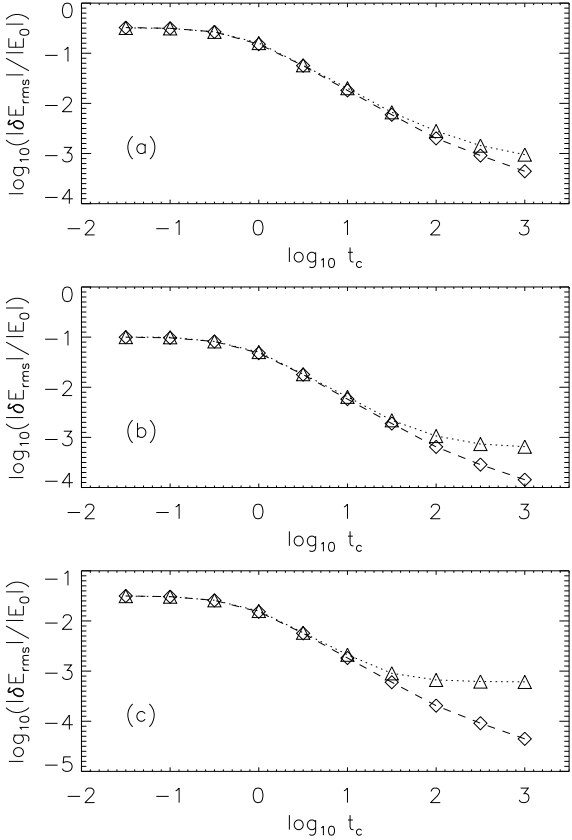}
           }
        \begin{minipage}{10cm}
        \end{minipage}
        \vskip -0.3in\hskip -0.0in
\caption{ 
(a) $|{\delta}E_{rms}|/|E_{0}|$, the fractional change in energy for 8000
orbit ensembles evolved in the dihedral potential with $E=4$, $a=b=1$,
and $D=10^{-4}$ after a time $t=256$, for variable $t_{c}$. Triangles
refer to an ensemble of regular initial conditions; diamonds refer to an
ensemble of chaotic initial conditions. (b) The same for $D=10^{-5}$.
(c) The same for $D=10^{-6}$.}
\vspace{0.0cm}
\label{landfig}
\end{figure}
\section{PHYSICAL IMPLICATIONS}
When applied to eq. (16), the expressions for $D$ and $t_{c}$ motivated in 
Section 2.3 allow one to relate the time scale associated with energy
relaxation to the dynamical time $t_{D}$. 

Consider first the effects of a near-random external environment where,
presumably, $t_{c}>t_{D}$. In this case, one infers that
\begin{displaymath}
\left({{\delta}E_{rms}\over |E|}\right)^{2}{\;}{\sim}{\;} 
\left({D\over |E|}\right)\left({t_{D}^{2}\over t_{c}^{2}}\right)t 
\end{displaymath}
\begin{equation}
{\;\;\;\;\;\;\;\;\;\;\;\;\;\;\;\;\;\;\;\;}
{\sim}{\;}\left({G^{2}M_{g}^{2}\over R_{g}^{4}|E|}\right) 
\left({R_{g}^{4}\over R_{s}^{4}}\right) 
\left({t_{D}^{2}\over t_{c}}\right) t
\end{equation}
Suppose, however, that the galaxy is at least roughly in virial equilibrium. In
this case, the `Virial Theorem' implies that
\begin{equation}GM_{g}/R_{g}{\;}{\sim}{\;}|E|
\end{equation}
and the dynamical time $t_{D}$ may be taken to satisfy
\begin{equation}
t_{D}^{2}{\;}{\sim}{\;}{1\over G{\rho}}{\;}{\sim}{\;}{R_{g}^{3}\over (GM_{g})}.
\end{equation}
Substitution of eqs (28) and (29) into eq. (27) then leads to the conclusion 
that
\begin{equation}
{{\delta}E_{rms}\over |E|}{\;}{\sim}{\;}
\left({R_{g}\over R_{s}}\right)^{2}
\left({t_{D}\over t_{c}}\right)^{1/2}\left({t\over t_{D}}\right)^{1/2}.
\end{equation}

In a similar fashion, one can allow for the effects of internal pulsations,
assuming, as in Section 2.3, that $t_{c}$ is less than or comparable to 
$t_{D}$. In this case, ${\delta}E_{rms}$ satisfies
\begin{equation}
\left({{\delta}E_{rms}\over |E|}\right)^{2}{\;}{\sim}{\;} 
\left({Dt\over |E|}\right){\;}{\sim}{\;}
{\alpha}^{2}{\beta} \left({GM_{g}\over R_{g}^{4}}\right)\, t_{D}\,t,
\end{equation}
and by again exploiting eqs. (28) and (29), one concludes that
\begin{equation}
{{\delta}E_{rms}\over |E|}{\;}{\sim}{\;}
{\alpha}{\beta}^{1/2}
\left( {t\over t_{D}} \right)^{1/2}.
\end{equation}

Equations (30) and (32) allow one to infer the typical degree of energy
relaxation that can be expected over a time scale ${\sim}{\;}t_{H}$, the
age of the Universe. 

Consider, {\it e.g.,} the effects of a dense cluster environment where, not 
unrealistically, the typical velocity dispersion associated with stars in an 
individual galaxy is comparable in magnitude to the velocity dispersion of 
galaxies moving in the cluster, and where the typical distance between 
galaxies is of order six times the size of a typical galaxy. In this case,
$R_{s}{\;}{\sim}{\;}6R_{g}$ and $t_{c}{\;}{\sim}{\;}6t_{D}$, so that, assuming
that $t_{H}{\;}{\sim}{\;}100t_{D}$, one would anticipate a fractional root 
mean squared energy ${\delta}E_{rms}/|E|{\;}{\sim}{\;}0.1$ over the lifetime
of the galaxy. Note that, as discussed in Section 2.3, this estimate likely
understimates by a small amount the efficacy of energy relaxation induced by 
random interactions with nearby galaxies.

Similarly, for the case of internal pulsations, it might seem plausible to
assume that the characteristic fractional amplitude of the perturbations 
${\alpha}{\;}{\sim}{\;}10^{-2}$ and that $t_{c}$ is comparable to $t_{D}$,
so that ${\beta}{\;}{\sim}{\;}1$. One would then anticipate once again that, 
within a time $t_{H}{\;}{\sim}{\;}100t_{D}$, the fractional root mean squared 
change in energy will have grown to a value 
${\delta}E_{rms}/|E|{\;}{\sim}{\;}0.1$.

These estimates effects -- changes in energy by of order 10\% over the age
of the Universe -- are not huge, but they could prove important in real
galaxies. Recent numerical work suggests that, because of the role of chaos
in certain key phase space regions, {\it e.g.}, near corotation and other 
resonances in spiral galaxies ({\it cf}. Wozniak 1993) or in the central 
regions of cuspy-core ellipticals ({\it cf}. Merritt \& Fridman 1996), it may 
be hard for a galaxy to settle down towards a true collisionless equilibrium. 
However, it would seem substantially easier ({\it cf}. Siopis \& Kandrup 2000) 
for the system to evolve towards a configuration which, albeit not 
corresponding to a true equilibrium, could persist in isolation as a 
near-equilibrium for times long compared with the age of the Universe. 

The important point, then, is that such 
near-equilibria, even if nearly stable as isolated entities over intervals
${\sim}{\;}t_{H}$, could be destabilised by low amplitude perturbations of
the form to which real galaxies are necessarily subjected. There is compelling
numerical evidence that chaotic orbits are very susceptible to perturbations
which induce phase space transport on (or very nearly on) the constant energy
hypersurfaces ({\it cf}. Kandrup, Pogorelov, \& Sideris 2000, Siopis \& 
Kandrup 
2000), thus facilitating extrinsic diffusion along the Arnold web. Such 
diffusion will occasion changes in the phase space density, which in turn 
alters the form of the gravitational potential, which could in principle 
triggering a nontrivial secular evolution. 

One might, however, argue that this need not have significant implications for
bulk evolution since most of the phase space is (presumably) dominated by
regular orbits which ({\it cf.} Binney 1978) support the skeleton of the 
system and which, significantly,
are not susceptible to such phase space diffusion. The key recognition,
therefore, is that energy relaxation impacts both regular and chaotic orbits;
and that even comparatively small effects acting on regular orbits, combined
with larger effects acting on the (presumably) smaller measure of chaotic
orbits, could effectively destabilise a near-equilibrium on a time scale
$<t_{H}$.

One plausible scenario might involve the effects of internal oscillations
triggered by a near-collision with another galaxy. Numerical simulations
({\it cf.} Vesperini \& Weinberg 2000) suggest that such near-collisions can 
serve to induce comparatively long lived, large amplitude oscillations which 
could, {\it e.g.}, account for the offset nuclei and/or lopsided and warped
discs observed in certain galaxies ({\it cf.} Kornreich, Haynes, \& Lovelace
1998 and references cited therein); and even if these oscillations are 
dominated by a small number of lower order normal modes, they should be 
accompanied by a larger number of higher order modes which could reasonably be 
modeled as near-random excitations.

In this context, it important to note that energy relaxation induced by
near-random internal irregularities is likely to be more important for the
bulk structure of a galaxy than more systematic, nearly periodic effects.
Periodic, or near-periodic, driving, associated with a small number of 
companion objects or with a small number of normal or pseudo-normal modes, 
involves perturbations with a countable set of discrete frequencies. The
obvious point, then, is that since these perturbations have their effect
because of a resonant coupling to orbits in the galaxy, they can only have 
a large effect on orbits which have considerable power at or near these
special frequencies and/or harmonics thereof. Chaotic orbits typically have
broad band Fourier spectra ({\it cf}. Tabor 1989) and, as such, can couple 
efficiently to such periodic disturbances ({\it cf}. Kandrup, Abernathy, \& 
Bradley 
1995). However, for regular orbits the power is concentrated at a few special 
frequencies and, if these frequencies are not in a resonance or near-resonance 
with the frequencies of the perturbations, the perturbation will have a 
comparatively minimal effect. Periodic driving could perhaps prove important 
for chaotic orbits in general or for regular orbits in specific phase space 
regions. However, one might anticipate generically that it will be less 
important on the overall phase space structure which, presumably, is dominated 
by regular orbits with frequencies that are not in resonance with the periodic 
driving.
\section*{Acknowledgments}
This research was supported in part by the National Science Foundation grant
NSF AST-0070809.
I am pleased to acknowledge useful discussions with Christos Siopis, Ioannis
Sideris, and, especially, Dick Miller and Alexi Fridman.

\label{lastpage}
\vfill\eject\end{document}
\end{document}